\documentclass[journal,transmag]{IEEEtran}
\usepackage{color}
\usepackage{ifpdf}

\usepackage{cite}

\ifCLASSINFOpdf
   \usepackage[pdftex]{graphicx}
\else
\fi
\usepackage{amsmath}
\usepackage{amssymb}

\newcommand{\units}[3][]{$#1\mathrm{#2\,#3}$}

\begin{document}

\title{Experimental investigation of the temperature-dependent magnon density and its influence on studies of spin-transfer-torque-driven systems}

\author{\IEEEauthorblockN{Thomas Meyer\IEEEauthorrefmark{1},
Thomas Br\"acher\IEEEauthorrefmark{1},
Frank Heussner\IEEEauthorrefmark{1}, 
Alexander A. Serga\IEEEauthorrefmark{1},
Hiroshi Naganuma\IEEEauthorrefmark{2},\\
Koki Mukaiyama\IEEEauthorrefmark{2},
Mikihiko Oogane\IEEEauthorrefmark{2},
Yasuo Ando\IEEEauthorrefmark{2},
Burkard Hillebrands\IEEEauthorrefmark{1}, and
Philipp Pirro\IEEEauthorrefmark{1}}

\IEEEauthorblockA{\IEEEauthorrefmark{1}Fachbereich Physik and Landesforschungszentrum OPTIMAS, Technische Universit\"at Kaiserslautern, \\67663 Kaiserslautern, Germany}
\IEEEauthorblockA{\IEEEauthorrefmark{2}Department of Applied Physics, Graduate School of Engineering, Tohoku University, Sendai 980-8579, Japan}
\thanks{Corresponding author: T. Meyer (email: thmeyer@physik.uni-kl.de).}}

%




\IEEEtitleabstractindextext{%

\begin{abstract} 
We present the temperature dependence of the thermal magnon density in a thin ferromagnetic layer. By employing Brillouin light scattering and varying the temperature, an increase of the magnon density accompanied by a lowering of the spin-wave frequency is observed with increasing temperature. The magnon density follows the temperature according to the Bose-Einstein distribution function which leads to an approximately linear dependency. In addition, the influence of this effect in spin-transfer-torque-driven systems is presented. In particular, the increase in the magnon density with temperature sets the limit for a suppression of magnons in charge current-driven systems. Hence, the maximum possible suppression of thermal magnons occurs at a finite current.
\end{abstract} 

}

\maketitle

\IEEEdisplaynontitleabstractindextext

%
\IEEEpeerreviewmaketitle
\section{Introduction}
%
%
%
%
\IEEEPARstart{T}{he} research field of magnon spintronics, where magnons, the quanta of spin-waves, are investigated receives a lot of interest. Many publications using magnons as information carrier are proof of principle studies aiming for future applications. Some examples for magnon-based logic devices are the magnon transistor~\cite{Chumak2014} and the magnon majority gate~\cite{Khitun2008,Fischer2017}. In addition, several ways to steer magnons in magnonic networks were developed, e.g., the spin-wave multiplexer~\cite{Vogt2014}, tunable spin-wave nanochannels~\cite{Wagner2016} or the steering of magnons via graded-index magnonics~\cite{Davies2015}.

But since the spin-wave damping limits the magnon lifetime and, thus, the spin-wave propagation length, the manipulation of the effective spin-wave damping is of particular interest for any future device using magnons. One way to control the effective spin-wave damping is the spin-transfer torque~(STT)~effect~\cite{Slonczewski1996} in combination with the spin-Hall effect~(SHE)~\cite{Hirsch1999}. If the measurement geometry is chosen adequately, this allows for a manipulation of the effective spin-wave damping by applying a charge current. This leads, e.g., to the realization of spin-Hall nano oscillators~\cite{Hamadeh2014,Demidov2012,Demidov2014,Lauer2017,Kajiwara2010,Collet2016} where a spin-wave auto-oscillation is observed if the spin-wave damping is compensated by the SHE-STT effect. However, in these charge current-driven systems, in general, also a temperature increase due to heating effects occurs. This can be observed by a frequency shift of the spin-wave spectrum since for increasing temperature, the saturation magnetization~$M_\text{Sat}(T)$ of a ferromagnetic layer decreases. 

A temperature change may also lead to the generation of a temperature gradient which is the basis for the research field of magnon caloritronics~\cite{Bauer2012}. In this field, the spin-Seebeck effect~(SSE)~\cite{Uchida2008} is employed to generate pure spin currents by a thermal gradient across the interface between a magnetic material and a nonmagnetic metal~\cite{Kehlberger2015}. It is reported that this spin current can manipulate the spin-wave damping in analogy to the SHE-STT effect~\cite{Jungfleisch2013,Lu2012}. However, not only a temperature gradient but also the absolute temperature influences the magnonic system~\cite{Langner2017}. This leads to a temperature-dependent spin-wave wave-vector for a given frequency~\cite{Obry2012} since the magnetization strongly influences the spin-wave dispersion. However, also an increase of the thermal magnon density for an increased temperature is expected from theory since magnons are distributed via the Bose-Einstein distribution~\cite{Demokritov2006} which is neglected in many studies.

In this contribution, we present the temperature-dependent magnon density measured by means of Brillouin light scattering microscopy~($\mu$BLS). This powerful technique allows for the detection of the frequency and the intensity of incoherent thermally excited spin waves. A more detailed description of the experimental technique can be found in~\cite{Sebastian2015}. The results reveal an increase of the magnon density with temperature which can be well described by the Bose-Einstein \mbox{(quasi-)particle} statistics. Furthermore, the effect of an increased temperature in a charge current-driven system via the SHE-STT effect is reported. In particular, we show that the suppression of the thermal magnon density due to an increased damping is limited by the Joule heating of the charge current.

 


\section{Temperature-dependent magnon density}
\subsection{Experimental results}
The investigated sample consists of a \units{5}{nm} thick Cr buffer layer which is deposited on a MgO substrate using a sputtering technique. On top, a \units{5}{nm} thick $\text{Co}_2\text{Mn}_{0.6}\text{Fe}_{0.4}\text{Si}$~(CMFS) layer is deposited. This low-damping ferromagnetic full Heusler compound~\cite{Sebastian2012,Sebastian2015-2} constitutes the magnetic layer of the stack and is capped by a Pt layer with \units{2}{nm} thickness. This layer allows for an efficient conversion of a charge current into a pure spin current for the SHE-STT-driven magnetization dynamics investigated in Sec.~\ref{Sec:STT}.\\
%
%
%
%
\begin{figure}[!t]
	 \centering
	 \includegraphics[width=0.4\textwidth]{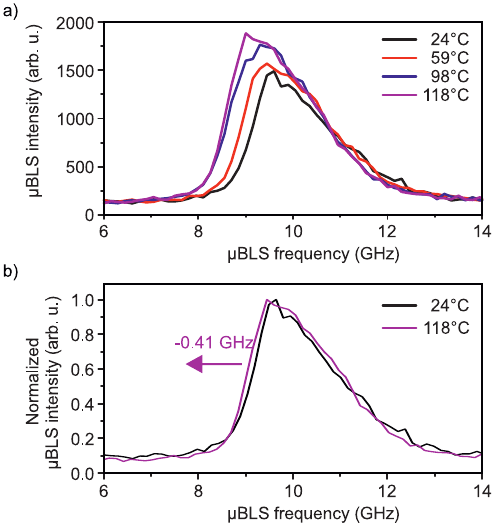}
	\caption{~a)~Experimentally obtained spin-wave spectra for different sample temperatures. For increasing temperature, an increase in the spin-wave intensity and a lowering of the spin-wave frequency is observed. b)~Normalized $\mu$BLS intensity for a sample temperature of \units{24}{^\circ C} (black line) and \units{118}{^\circ C} (pink line). The latter one is shifted in frequency by \units{-0.41}{GHz} resulting in an overlapping of both spectra and the identical shape can clearly be seen.}
	\label{Fig:Spectra}
\end{figure}
For the $\mu$BLS measurements in this Section, the sample temperature is increased from room temperature~$T_\text{R}=24~^\circ \text{C}$ to \units{118}{^\circ C} by using a heating element underneath the sample. Since the thickness of the waveguide is small and all layers are metallic, they are thermally well-coupled. Thus, the temperature gradients between the magnetic layer and the adjacent non-magnetic metallic layers are negligible resulting in a vanishing influence of the SSE in the presented measurements. Figure~\ref{Fig:Spectra}a exemplarily shows the obtained temperature-dependent thermal spin-wave spectra for an externally applied magnetic field of $\mu_0 \text{H}_\text{ext}=70~\text{mT}$. Since the measured $\mu$BLS intensity is proportional to the magnon density, the results reveal a significant increase of the magnon density with increasing temperature and a lowering of the spin-wave frequency spectrum.

The frequency-dependent $\mu$BLS intensity is given by the wave-vector-dependent detection efficiency of the setup~$\eta_{\mu \text{BLS}}(\textbf{k})$ and the spin-wave density of states~$D(\textbf{k})$. The latter depends on the spin-wave dispersion relation~$f (\textbf{k},\phi, \text{M}_\text{S}, \text{H}_\text{ext})$ which also takes the external magnetic field~$\text{H}_\text{ext}$ and the angle of the propagation~$\phi$ with respect to the magnetization~$\textbf{M}$ (and the saturation magnetization $\text{M}_\text{S}$) into account. In the used setup, $\eta_{\mu \text{BLS}}(\textbf{k})$ vanishes for an in-plane spin-wave wave vector $\textbf{k}_{\vert \vert} \gtrsim 17~\text{rad}/\mu \text{m}$~\cite{Sebastian2015}. 

Additionally, also the energy- and temperature-dependent magnon distribution function~$N(f,T_\text{S})$ needs to be taken into account. This yields a frequency- and temperature-dependent $\mu$BLS intensity of:
%
%
%
%
\begin{equation}
		I_{\mu \text{BLS}} (f,T_\text{S}) \propto \sum_{\textbf{k},\atop f (\textbf{k})=f} \eta_{\mu \text{BLS}}(\textbf{k}) D(\textbf{k}) N(f,T_\text{S}).
			\label{Eq:BLS-Int}
\end{equation}

Due to the anisotropy of $D(\textbf{k})$ with respect of the propagation angle of the spin waves with respect to {\bf M}, and the decrease of $\eta_{\mu \text{BLS}}(\textbf{k})$ with increasing {\bf k}, the term $\eta_{\mu \text{BLS}}(\textbf{k}) D(\textbf{k})$ leads to an asymmetric shape of the spectra~\cite{Sebastian2015}.

It is important to note that, for the measurements presented in this Letter, this shape does not change on increasing the temperature but the spectra are only slightly shifted in frequency. Hence, the spectra can be expressed by a function which depends solely on the frequency relative to a temperature-dependent center frequency: $I_{\mu \text{BLS}} (f-f_\text{Center}(T_\text{S}))$.

To demonstrate this, Fig.~\ref{Fig:Spectra}b shows the obtained $\mu$BLS intensity for a sample temperature of $T_\text{S}= T_\text{R} = 24^\circ\text{C}$ (black line) and $T_\text{S}= 118^\circ\text{C}$ (pink line) normalized to the maximum detected intensity and shifted by $\Delta f = -0.41~\text{GHz}$. In this case, both spectra overlap which reveals their identical shape. Thus, the increase of the $\mu$BLS intensity with increasing temperature is solely determined by $N(f_\text{Center}, T_\text{S})$:
%
%
%
%
\begin{equation}
		I_{\mu \text{BLS}} (f,T_\text{S}) \propto N(f_\text{Center}, T_\text{S}) \cdot I_{\mu \text{BLS}} (f-f_\text{Center}(T_\text{S})).
			\label{Eq:BLS-Intsplit}
\end{equation}

To allow for a detailed investigation of the underlying effects, in the following, the intensity increase and the frequency shift for increased $T_\text{S}$ are determined from the measurements.

The latter is expected from a reduction of~$M_\text{Sat}(T)$ with increasing temperature and a subsequent shift of the spin-wave dispersion~\cite{Kalinikos1986,Braecher2017}. However, to quantitatively determine the frequency-shift, a reliable method to evaluate the measured spin-wave frequencies is needed. Since many spin-wave modes contribute to the measured data, a peak fit of the spectra is not applicable to determine the frequency shift. Thus, in this Letter, $f_\text{Center}$ is determined as the weighted average of the measured frequency-dependent intensity~$I_{\mu \text{BLS}}(f)$:

\begin{equation}
\label{Eq:fcenter}
f_\text{Center}=\sum I_{\mu \text{BLS}}(f) f / \sum I_{\mu \text{BLS}} (f)
\end{equation}

This method has several advantages since it does not rely on any assumptions like, e.g., the shape of the spectra.

The resulting relative frequency shift as a function of the increased sample temperature $\Delta f_\text{Center}(\Delta T)$ with respect to room temperature $\Delta T = T_\text{S}-T_\text{R}$ is shown by the black squares in Fig.~\ref{Fig:BLS-Temperature}a. The obtained frequency shift corresponds well to the frequency shift of \units{-0.41}{GHz} for the maximum achieved temperature increase of \units{96}{K} as indicated in Fig.~\ref{Fig:Spectra}b. The frequency error in Fig.~\ref{Fig:BLS-Temperature}a is given by the frequency resolution of the setup. Here, $\Delta f_\text{Center}(\Delta T)$ exhibits a linear slope and the red line depicts the according linear fit to the data yielding a temperature-dependent frequency shift of $\Delta f_\text{Center}/\Delta T= (-5.3 \pm 0.4)~\text{MHz/K}$. For small $\Delta T$, the changes in $f_\text{Center}$ are below the frequency resolution of the experimental setup. Thus, the linear fit is only applied to the data points at $\Delta T > 20 ^\circ \text{C}$.

\subsection{Increase of the magnon density with temperature}

\begin{figure}[!t]
	 \centering
	 \includegraphics[width=0.4\textwidth]{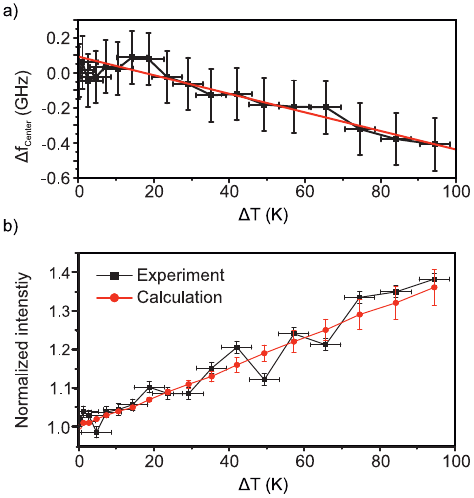}
	\caption{~a)~Shift of $f_\text{Center}$ determined from the measured spectra for increasing $\Delta T$ (black squares). The red line denotes a linear fit to the data. b)~Experimentally observed intensity increase with increasing $\Delta T$ normalized to the intensity at $T_\text{R}$ (black squares). The red circles show the calculated intensity increase via Eq.~\ref{Eq:RJ}, normalized to the calculated value at $T_\text{R}$.}
	\label{Fig:BLS-Temperature}
\end{figure}

To quantify the observed increase of the magnon density, the temperature-dependent $\mu$BLS intensity of the measured spin-wave spectra is integrated over all observed spin-wave frequencies and normalized to the $\mu$BLS intensity measured at $T_\text{R}$. The results are depicted by the black squares in Fig.~\ref{Fig:BLS-Temperature}b and show an intensity increase by a factor of up to 1.37.

In the previous Section, it is shown that the temperature-dependent spin-wave spectra obtained via $\mu$BLS can be described by a function which yields the shape of the spectra~$I_{\mu \text{BLS}} (f-f_\text{Center}(T_\text{S}))$ and the magnon distribution function~$N(f,T_\text{S})$ (Eq.~\ref{Eq:BLS-Intsplit}).

In general, $N(f,T_\text{S})$ for magnons is given by the Bose-Einstein distribution function for a vanishing chemical potential~\cite{Demokritov2006}:
%
%
%
%
\begin{equation}
 N(f,T_\text{S}) = \frac{1}{\text{e}^{\frac{\text{h}f}{\text{k}T_\text{S}}}-1}.
	\label{Eq:Bose}
\end{equation}
Here, k and $T_\text{S}$ denote the Boltzmann factor and the temperature, respectively, and h denotes Planck's constant. In the presented case, the assumption of a vanishing chemical potential is appropriate since the system is in thermal equilibrium with the lattice.
%
%
%
%

Considering that the accessible magnons in this experiment exhibit energies which correspond to the thermal energy at $T \leq 60~\text{mK}$, in this low-energy limit, the first-order approximation of the magnon density is given by the Rayleigh-Jeans distribution:
%
%
%
%
\begin{equation}
	  N(f,T_\text{S}) = \frac{\text{k}T_\text{S}}{\text{h}f}.
	\label{Eq:RJ}
\end{equation}
%
%
 
%
%
%
%
\begin{figure}[!t]
	 \centering
	 \includegraphics[width=0.4\textwidth]{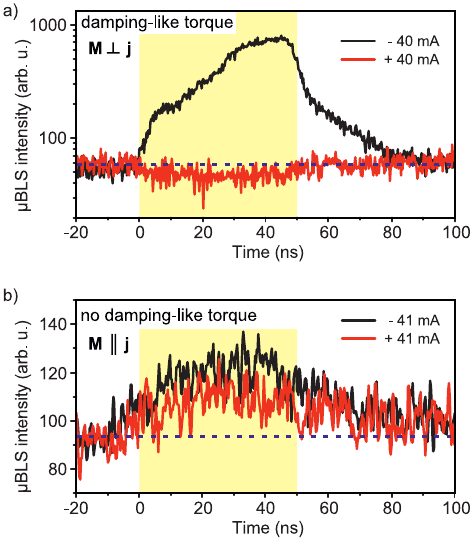}
	\caption{~a)~Time-dependent $\mu$BLS intensity (note the log-scale) for \mbox{$j=-40~\text{mA}$} (black line) and \mbox{$j=+40~\text{mA}$} (red line) in the Damon-Eshbach geometry. For $j<0$ ($j>0$), the spin-wave damping is decreased (enhanced) resulting in a increase (decrease) of the magnon density. b)~Time-dependent $\mu$BLS intensity (note the linear scale) for different currents in the Backward-volume geometry. The intensity increases for both current directions. The dashed line in both graphs indicates the noise level and the shaded area depicts the temporal extent of the current pulse.}
	\label{Fig:Timetraces}
\end{figure}

To determine the expected intensity increase in the experiments by Eq.~\ref{Eq:RJ}, $T_\text{S}$ and the shift of the spin-wave frequency need to be considered. For the investigation presented in this Letter, the latter is rather small $\Delta f_\text{Center} < 0.5~\text{GHz}$ compared to the absolute frequency of $8~\text{GHz} \lesssim f \lesssim 12~\text{GHz}$.

Hence, assuming a constant spin-wave frequency, Eq.~\ref{Eq:RJ} yields a maximum relative change of:
\begin{equation}
	\Delta N_\text{max} \mathrel{\mathop:}= \frac{N(f,T_\text{S,max})}{N(f,T_\text{R})} = \frac{T_\text{S,max}}{T_\text{R}} \approx 1.32.
	\label{Eq:DeltaN}
\end{equation}
This value is already in a good agreement with the experimental findings shown by the black squares in Fig.~\ref{Fig:BLS-Temperature}b.

Additionally, if also the temperature-dependent center frequency~$f_\text{Center}(T_\text{S}) = f_\text{Center}(T_\text{R}) + \Delta f_\text{Center}/\Delta T \cdot T_\text{S}$ is taken into account, Eq.~\ref{Eq:DeltaN} is modified to:
\begin{equation}
	\Delta N = \frac{T_\text{S}}{T_\text{R}} \cdot \frac{f_\text{Center}(T_\text{R})}{f_\text{Center}(T_\text{R})+\Delta f_\text{Center}/\Delta T \cdot T_\text{S}}.
	\label{Eq:DeltaN-frequency}
\end{equation}

According to this, the modeled increase of the $\mu$BLS intensity is depicted by the red circles in Fig.~\ref{Fig:BLS-Temperature}b. The error is mainly given by the deviations in the determination of $f_\text{Center}(T_\text{S})$. The values are in excellent agreement with the experimentally observed increase of the magnon density (black squares) and yield a maximum expected increase in the magnon density by a factor of 1.36.

In summary, the modeled increase of the $\mu$BLS intensity in this Section shows that the observed increase in the $\mu$BLS intensity is mainly caused by the increased temperature and the according increase of the Bose-Einstein distribution function. Furthermore, if the temperature-dependent frequency shift is taken into account, this yields a small additional increase of the modeled $\mu$BLS intensity which leads to an excellent agreement with the experimental data.

%
%
%
\begin{figure}[!t]
	 \centering
	 \includegraphics[width=0.43\textwidth]{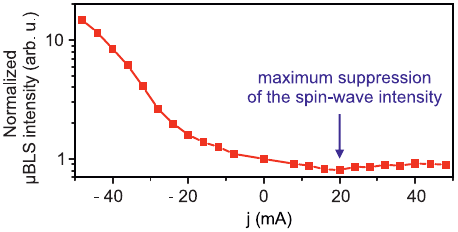}
	\caption{~Experimentally obtained $\mu$BLS intensity for the Damon-Eshbach geometry, integrated over all spin-wave frequencies and normalized to zero current (note the log-scale). The maximum suppression of the spin-wave intensity is obtained for an applied current of approximately $j=+20~\text{mA}$ as indicated by the blue arrow.}
	\label{Fig:IntegratedIntensity}
\end{figure}

\section{Temperature-induced effects in spin-transfer-torque-driven systems}
\label{Sec:STT}
To demonstrate the influence of this effect in SHE-STT effect driven systems, the layer stack is patterned into \units{(7 x 30)}{\mu m^2} large spin-wave waveguides. At each end of the waveguides, \units{300}{nm} thick Au contacts are deposited. These contacts allow for the application of a charge current~\textbf{\textit{j}} to the waveguide which will flow through all three layers of the layer stack since all layers are metallic. 

On the one hand, \textbf{\textit{j}} leads to a temperature increase due to Joule heating. On the other hand, the SHE in the Pt layer and in the Cr layer causes a partial conversion of the charge current into a pure spin current. Due to the opposite spin-Hall angles of Pt and Cr~\cite{Du2014,Schreier2014} and the fact that the spin currents enter the CMFS from opposite interfaces, the spin currents add up. Depending on the orientation of the external field with respect to \textbf{\textit{j}}, the torque on the magnetization~\textbf{\textit{M}} via the STT effect can be co-aligned with the Gilbert damping torque. 

Since \textbf{\textit{j}} and \textbf{\textit{M}} are always aligned in-plane, in the case of $\textbf{\textit{M}} \perp \textbf{\textit{j}}$, hence, $\textbf{\textit{M}}$ pointing along the short axis of the waveguide, the effective spin-wave damping is manipulated by $j$. This geometry is usually also referred to as Damon-Eshbach geometry~\cite{Eshbach1960}. In the following, \units{50}{ns} long current pulses are applied. Thus, in the case of a reduced damping, an increase of the magnon density is observed by means of time-resolved $\mu$BLS as depicted by the black line in Fig.~\ref{Fig:Timetraces}a in the case of $j < 0 $ and a magnetic field of \units{70}{mT}. Since the threshold current of approximately \units{-16}{mA} is overcome, which refers to the current at which the spin-wave damping is completely compensated, the intensity increases exponentially in time until the end of the current pulse. In contrast, for $j > 0 $, the STT effect increases the effective spin-wave damping. This results in a decrease of the magnon density below the initial thermal level (dashed line) in the absence of a current as can be seen by the red line in Fig.~\ref{Fig:Timetraces}a.

In order to distinguish the influence of Joule heating on the spin-wave intensity and the influence of the STT effect, the measurement geometry is changed to the so-called backward-volume geometry. In this case, $\textbf{\textit{M}}$ is aligned along the long axis of the waveguide, hence, $\textbf{\textit{M}}~\vert \vert~\textbf{\textit{j}}$. This geometry does not allow for a damping manipulation. Thus, the spin-wave damping remains unchanged during the current pulse. 

However, as presented in Fig.~\ref{Fig:Timetraces}b, still an intensity increase during the current pulse is observed. Furthermore, in contrast to the previous results, the intensity increase appears for both current directions. Thus, the observed increase originates from an enhanced temperature of the waveguide due to Joule heating by the charge current. 

Comparing the observed intensity increase of a factor of approximately 1.22 to the previously measured and calculated temperature-dependent intensity increase (Fig.~\ref{Fig:BLS-Temperature}b), this yields a temperature increase of \units{(60 \pm 20)}{K}. It should be noted that this value constitutes a lower estimation of the heating due to the finite noise level in this measurement.

Even though this effect seems to be rather small, it strongly affects the suppression of the magnon density in the case of a damping enhancement in the Damon-Eshbach geometry. To demonstrate this, Fig.~\ref{Fig:IntegratedIntensity} shows the measured time- and frequency-integrated $\mu$BLS intensity in the case of the Damon-Eshbach geometry (note the log-scale) during the current pulses, normalized to zero current. The large intensity increase for $j<0$ by up to a factor of 15 due to the STT effect can clearly be seen. In contrast, for $j > 0$, the spin-wave damping is enhanced leading to a suppression of the thermal magnon density. Here, the maximum suppression is found to occur not for the maximum current of $j=+48~\text{mA}$ but for a current of $j \approx +20~\text{mA}$ as indicated by the blue arrow. In combination with the previously demonstrated results of this Letter, this indicates a temperature increase due to Joule heating which results in an overall increase in the magnon density. Since the STT effect scales linearly with $j$ and the Joule heating is proportional to $j^2$, this causes the partial compensation of the STT effect-induced suppression of the magnon density for large positive currents. This effect needs to be taken into account in any future application using magnons and in which the thermal magnon density is suppressed via the STT effect.

\section{Conclusion}
In conclusion, we have experimentally investigated the temperature-dependence of the magnon density. The results show the increase of the overall magnon density due to an increased temperature. By taking the shift of the spin-wave dispersion due to a lowered saturation magnetization into account, the results can be well reproduced by the Bose-Einstein distribution of magnons without the consideration of a spin current. This effect is usually not taken into account in SSE-driven measurements since the investigations only consider a temperature gradient but neglect changes in the absolute temperature. In this Letter, we could also show the influence of this effect in SHE-STT effect-driven systems. By changing the measurement geometry, we could verify that the temperature of the spin-wave waveguide increases due to Joule heating which, in return, leads to an intensity increase which can be modeled without taking a temperature gradient into account. This effect also strongly limits the possible suppression of thermally excited magnons by the STT effect in the investigated system.

\section*{Acknowledgment}
The authors gratefully acknowledge financial support by the DFG in the framework of the Research Unit TRR 173 ``Spin+X'' (Project B01), and by the DFG Research Unit 1464 and the Strategic Japanese-German Joint Research Program from JST: ASPIMATT.

\ifCLASSOPTIONcaptionsoff
  \newpage
\fi



%

\end{document}